\documentclass[preprint]{aastex}
\shorttitle{Constraints on Dark Energy}
\shortauthors{Daly \& Djorgovski}
\def\lsim{\hbox{ \rlap{\raise 0.425ex\hbox{$<$}}\lower 0.65ex\hbox{$\sim$} }}
\def\gsim{\hbox{ \rlap{\raise 0.425ex\hbox{$>$}}\lower 0.65ex\hbox{$\sim$} }}

\begin{document}

\title{A Model-Independent Determination of the Expansion and Acceleration Rates
of the Universe as a Function of Redshift and Constraints on Dark Energy}

\author{Ruth A. Daly}

\affil{Department of Physics, Berks-Lehigh Valley College,
Pennsylvania State University, Reading, PA, 19610}
\email{rdaly@psu.edu}

\and

\author{S. G. Djorgovski}

\affil{Division of Physics, Mathematics, and Astronomy,
California Institute of Technology, MS 105-24, Pasadena, CA 91125}
\email{george@astro.caltech.edu}

\begin{abstract}
Determination of the expansion and acceleration history of the universe
is one of the fundamental goals of cosmology.  Detailed measurements
of these rates as a function of redshift can provide new physical insights
into the nature and evolution of the dark energy, which apparently dominates
the global dynamics of the universe at the present epoch.
We present here dimensionless coordinate distances $y(z)$ to twenty radio
galaxies reaching out to $z \approx 1.8$, the redshift range currently not
covered by Supernova standard candle observations.  There is very good
agreement between coordinate distances to radio galaxies and supernovae
for the redshift range where these measurements overlap, suggesting
that neither is plagued at this level by unknown systematic errors.  
We develop a simple numerical method for a direct determination of the
expansion and acceleration rates, $E(z)$ and $q(z)$, from the data,
which makes no assumptions about the underlying cosmological model or
the equation of state parameter $w$.  This differential method is in contrast
the traditional cosmological tests, where particular model equations are
integrated and then compared with the observations.  The new approach is
model-independent, but at a cost of being noisier and highly sensitive to
the amount and quality of the available data.
We illustrate the method by applying it to the currently available Supernova
data and the data on radio galaxies presented here.  We derive the expansion
rate of the universe as a function of redshift, $E(z)$, and for the first
time obtain 
a direct estimate of the acceleration rate of the universe as a function
of redshift, $q(z)$, in a way that is independent of assumptions regarding
the dark energy and its redshift evolution. 
The current observations 
indicate that the universe transitions from acceleration to deceleration
at a redshift greater than 0.3, with a best fit estimate of about 0.45;
this transition redshift and our determinations of $E(z)$ are  
broadly in agreement with the currently popular Friedmann-Lemaitre cosmology
with $\Omega_m = 0.3$, and $\Omega_{\Lambda} = 0.7$, even though no model
assumptions are made in deriving the fits for $E(z)$ and $q(z)$.
With the advent of much better and
richer data sets in the future, our direct method can provide a useful
complementarity and an independent check to the traditional cosmological tests.

\end{abstract}


\section{Introduction}

A traditional task of cosmology is to determine the global geometry and
dynamics of the universe.  The field has been revolutionized by the modern
measurements of CMBR fluctuations
(e.g., Bennett et al. 2003, Spergel et al. 2003, and references therein),
the use of distant supernov\ae\ (SNe) in a Hubble diagram
(see, e.g., Riess 2000, Leibundgut 2001, and references therein), 
radio galaxies (e.g. Daly \& Guerra 2002), 
and many other advances.  What these modern measurements have now convincingly
demonstrated is that the global mass/energy budget of the universe, and thus
its dynamics, is dominated by a so-called dark energy component, which accounts
for $\gsim 70$\% of the closure density today.  Einstein's cosmological
constant, $\Lambda$, is one special (and viable) case.  More generally, this
mysterious dark energy component is characterized through the equation of
state, $w = p/\rho$, where $p$ is the pressure and $\rho$ the energy density;
the cosmological constant solution corresponds to $w = -1$.  
For reviews and further references, see, for example, 
Sahni \& Starobinsky (2000), Turner (2002a, 2002b),
Peebles \& Ratra (2003), and 
Padmanabhan (2003).

The nature of the dark energy (including its evolution in redshift, if any)
is one of the most outstanding problems of physics and astronomy today.
Constraining it through analysis of cosmological data is a task of a critical
importance, and every new data set or analysis method can provide valuable
insights into this problem.

Several recent studies have focused on the use of 
supernovae to determine the properties of the dark
energy (Starobinsky 1998; Huterer \& Turner 1999;
Saini et al. 2000; Chiba \&
Nakamura 2000; Maor, Brustein, \& Steinhardt 2001; 
Goliath et al. 2001; Astier 2001; 
Gerke \& Efstathiou 2002; Weller \& Albrecht 2002; 
and Padmanabhan \& Choudhury 2002).
The key ingredients are luminosity distances 
to sources over a broad
range of redshift, preferably including sources at
high redshift. 
Most of these analyses have focused on constraints
on an evolving scalar field such as that used to 
define quintessence (Caldwell, Dave, \& Steinhardt) or
a rolling scalar field (Peebles \& Ratra 1988).    
More recently, other types of models have been proposed
to account for the acceleration of the universe, such
as stringy dark energy (Frampton 2002)
and k-essence (Armendariz-Picon, Damour, \& Mukhanov
1999; Barger \& Marfatia 2001).

Here, we focus on direct empirical determinations of the dimensionless
expansion rate $E(z)$ and acceleration rate $q(z)$ as functions of
redshift.  These require values for dimensionless coordinate distances to
sources over a broad range of redshifts.  We provide both a new data set,
and a new method for estimating of $E(z)$ and $q(z)$.

We first present coordinate distances to 20 radio galaxies (RGs), reaching
out to $z \approx 1.8$, and thus supplementing the existing SN data in what
is a critical redshift regime.  These RG data can be used to compare model
predictions of any flavor of dark energy with the observations.  These
coordinate distances are derived and listed in \S 2.  For completeness, and to
compare the RG and supernova SN results, the coordinate distances to 78 SNe
are also listed in \S 2.
In \S 3, we derive the expressions for a direct determination of $E(z)$ and
$q(z)$ from measurements of the dimensionless coordinate distances $y(z)$.
In \S 4, we describe our simple numerical differentiation technique which can
be used to implement these concepts on the real data.
We illustrate the method and present our preliminary results based on the
current RG and SN data sets in \S 5, and discuss implications for
the properties of the dark energy in \S 6.     
A summary and discussion follows in \S 7.

\section{Dimensionless Coordinate Distances}

The values of coordinate distances to sources at high redshift
can be used to determine or constrain global cosmological parameters,
and to understand the properties and redshift evolution of the dark energy.  
Coordinate distances $(a_or)$ may be obtained from luminosity distances $d_L$
or angular size distances $d_A$, since these are simply related to
the coordinate distance:
$d_L = (a_or)(1+z)$, and $d_A = (a_or)/(1+z)$ (e.g. Weinberg 1972).

The dimensionless coordinate distance, $y(z)$ is simply related to the 
coordinate distance $a_or$, $y(z) = H_0(a_or)$ (e.g., 
Carroll, Press, \& Turner 1992; Peebles 1993).  
The luminosity distance $d_L$ and the angular size distance
$d_A$ are also simply related to the dimensionless coordinate
distance:
$d_L = H_0^{-1}~y(z)~(1+z)= H_0^{-1}D_L$,
where $D_L$ is the dimensionless luminosity distance (e.g. 
Perlmutter et al. 1999), and
$d_A = H_0^{-1}~y(z)/(1+z)$.

Observations of type Ia supernovae and type IIb radio galaxies 
allow estimates of the 
dimensionless coordinate distances to sources at different redshift.

The use of FRIIb radio galaxies to determine the angular size 
distance or coordinate distance 
to radio galaxies at different redshifts 
is described in detail elsewhere (e.g. Podariu et al. 
2003; Daly \& Guerra 2002; Guerra, Daly, \& Wan 2000; Daly 1994).
In addition to the use of FRIIb radio galaxies addressed here,
other methods of using radio galaxies and quasars
to determine coordinate distances  
are discussed by Buchalter et al. (1998), 
Gurvits, Kellermann, \& Frey (1999), Vishwakarma (2001),
Lima \& Alcaniz (2002), and Chen \& Ratra (2003).  
Here FRIIb radio galaxies are used to obtain dimensionless
coordinate distances to 20 radio galaxies following the
method described, for example, by Daly \& Guerra (2002).

In the radio galaxy method proposed by Daly (1994),   
one model parameter $\beta$ enters into the
ratio $R_* \equiv <D>/D_*$; this ratio also depends on observed
quantities and the dimensionless coordinate distance $y(z)$.  In
this model the ratio $R_*$ is equal to a constant, $\kappa$:
\begin{equation}
R_* (\beta, y(z)) = \kappa~.
\end{equation}
The constants $\kappa$ and $\beta$ and their uncertainties
are obtained by fitting all of the data to equation (1),
as described in detail by Guerra, Daly, \& Wan
(2000), and Daly \& Guerra (2002).  The ratio is given by 
\begin{equation}
R_* = k_o y^{(6 \beta -1)/7}~(k_1y^{-4/7}+k_2)^{(\beta/3-1)}~,
\end{equation}
where $k_o$, $k_1$, and $k_2$ are observed quantities (described
in detail in the Appendix of Guerra, Daly, \& Wan 2000).
Equation (1) with $R_*$ given by equation (2) allows
a determination of $y(z)$ to each source; $y(z)$ is
implicitly known for each source and is determined using 
an iterative technique.  The values of $y$ obtained along
with the one $\sigma$ error of y are listed in 
Table 1.  In determining the
one $\sigma$ error bar on $y(z)$, the uncertainties of $\kappa$, 
$k_o$, $k_1$, and $\beta$ have been included; 
$k_2$ is known to high precision as it is 
the energy density of the microwave background
radiation at the source redshift, and is the term that describes
the effects of inverse Compton cooling of relativistic electrons
by the microwave background radiation. 
The best fit
values of $\kappa$ and $\beta$ vary slightly depending upon 
whether just the radio galaxy data are fit, or whether both
the radio galaxy and supernova data are fit.  Values of
$y$ obtained using the best fit parameters to radio galaxies
alone are labeled $y_s$ in Table 1, and those obtained 
using
the best fit to both the radio galaxy and supernova data
are labeled $y_j$.  That is, the best
fit values of $\kappa$ and $\beta$ change slightly depending
upon whether just the radio galaxy data are fit 
(referred to with a subscipt ``s''), or whether
the radio galaxy and supernovae data are fit simultaneously
(referred to with a subscript ``j'').  
Best fit values of $\kappa$ and $\beta$ are listed in Table 2.
Note that the radio galaxy method
does not rely upon a low-redshift normalization; the best
fit values of $\kappa$ and $\beta$ are determined using all of the data.

The best fit values of $\kappa$, $\beta$, and ${\cal M_B}$ (described below) 
and their 
error bars are included in Table 2, where the 54 supernovae
included in the ``primary fit C'' of Perlmutter et al. (1999)
and the 20 radio galaxies discussed here were studied.  
Values obtained 
from the fits of Daly \& Guerra (2002) that allow for 
quintessence in a spatially flat
universe with separate (s) and joint (j) fits to the 
radio galaxy and supernovae data are
labelled ``Q.''  Best fit values obtained in the 
rolling scalar field model analyzed by Podariu et al. (2003)
are labelled ``SF.''  
As the number of data
points in the fit increases, the value of each constant, 
$\kappa$, $\beta$, and ${\cal M_B}$, 
becomes independent of the assumptions of the fit.  For
example, the 54 supernovae points yield a consistent value
of ${\cal M_B}$ for fits that include supernovae only, or radio 
galaxies and supernovae, and in a universe with quintessence
or a rolling scalar field.  Since the value of ${\cal M_B}$ changes
so little when fit in different models, and when fit including
or excluding radio galaxies, only one value of $y$ is listed for 
each supernova.  The 20 radio galaxy points 
show some small variations in the values of $\kappa$ and $\beta$
obtained with radio galaxies alone, or radio galaxies
and supernovae, obtained in a universe with quintessence or
a rolling scalar field.  As more radio galaxy data points
are added, the values of the constants will be more
accurately determined. New runs were done that include the 
full 78 supernovae listed here and the 
best fit parameters and their error bars are the same as
those listed in Table 2.  
      
The coordinate distances to the
supernovae are determined following
the procedures of Perlmutter et
al. (1999) and Riess et al. (1998).  In the application of supernovae
type Ia as a distance indicator, there is one model
parameter $\alpha$ which is used to determine the effective
apparent B band magnitude at maximum brightness
$m_B^{eff}$.  This is related to the dimensionless
coordinate distance $y(z)$:
\begin{equation}
m_B^{eff}(\alpha) = {\cal M_B} ~+~5~log[(1+z)~y(z)]~.
\end{equation}
The constant ${\cal M_B}$ is determined by fitting all of
the supernova data, and is simply related to the absolute 
magnitude of the supernova peak brightness $M_B$:
${\cal M_B} = M_B +25 -5~log~H_0$ (see Perlmutter et al.
1999).  Equation (3) is then used to determine $y(z)$ to 
each of the 54 supernovae in the ``primary fit C'' of
Perlmutter et al. (1999), the 37 supernovae
presented by Riess et al. (1998), and the 1 high-redshift supernova
published by Reiss et al. (2001), with the magnitude
of this source corrected for gravitational lensing 
(Benitez et al. 2002). The one $\sigma$ uncertainty 
of y is obtained by combining the 
uncertainties of ${\cal M_B}$ and
$m_B^{eff}$.  These values of listed in Table 2.  There are
14 sources that are present in both the Riess et al. (1998) and
Perlmutter et al. (1999) samples used here.  In the 
determinations of $E(z)$ and $q(z)$, average values of $y$ with 
appropriate error bars  
were used for these duplicate sources; these values are listed in
Table 4.    
The values of $y(z)$ are shown in Figures 1 and 2.  The good
agreement between coordinate distances determined using radio 
galaxies and supernovae at similar redshifts is easy to see in these figures.

To test the reliability of the values of $y(z)$ obtained for the 
radio galaxies, a comparison was made between cosmological
parameters obtained directly from the radio galaxy data alone and
those obtained from the values of $y_s$ listed in Table 1 in
a quintessence model (see line 1 of Table 2). 
Each value of $y_s$ was substituted into equation (3)
to obtain an equivalent effective apparent magnitude for the radio 
galaxy; the value of ${\cal M_B}$ obtained for supernovae alone in
a universe with quintessence (line 1 of Table 2) was adopted.  
These effective apparent magnitudes were then analyzed in a
universe with quintessence, and the best fit parameters and
their one sigma ranges compared with those obtained directly
from the radio galaxy data.  First, the $\chi^2$ per degree of
freedom went from 16.5/16 to 15.6/15, so the reduced $\chi^2$
remains fairly constant; the number of degrees of freedom
drops by one in the new fit since one new parameter, ${\cal M_B}$
is fit.  The one sigma range of $\Omega_m$ is 0.0 to 0.24 in
the original fit, and is 0.0 to 0.17 in the new fit.  The one
sigma range of $\Omega_Q$ is 0.76 to 1.0 in the original fit,
and is 0.83 to 1.0 in the new fit.  The one sigma range of 
$w$ in the original fit is -1.3 to -0.43 centered on -0.73,
and is -1.5 to -0.56 centered on -0.8 in the new fit.  The best
fit value of ${\cal M_B}$ is 23.83 $\pm 0.08$, compared with 
the input value ${\cal M_B}$ of 23.91 $\pm 0.03$ used to
define an effective apparent magnitude for each radio galaxy.
Thus, the cosmological parameters obtained directly from the
radio galaxies are very similar to those obtained from the
values of $y(z)$ listed in Table 1.

\section{Computation of E(z) and q(z) from the Coordinate Distances}

The determinations of the dimensionless coordinate distances
do not require any assumptions regarding cosmological parameters,
the dark energy, or the redshift evolution of these components once
the values of the constants $\kappa$, $\beta$, and ${\cal M_B}$
have been determined.  
The first and second derivatives of the 
dimensionless coordinate distance with respect to redshift 
can be used to construct a model-independent determination of 
the dimensionless expansion rate $E(z) = H(z)/H_0$, and
the acceleration rate $q(z) = -\ddot{a} a/\dot{a}^2$.

These follow from the relation between redshift $z$ and
the cosmic scale factor $a(t)$, $(a(t)/a_o) = (1+z)^{-1}$, and
the Robertson-Walker line
element, which describes a homogeneous isotropic expanding universe, 
\begin{equation}
{d \tau^2 = dt^2 - a^2(t) 
\left( {dr^2 \over 1-kr^2} +r^2 d\theta^2 + r^2~sin^2\theta ~d\phi^2 \right)}
\end{equation}
(see, for example, Weinberg 1972).  It is well known
that these imply  
$H(z) \equiv \dot{a}/a = \sqrt{1-kr^2}~~(dy/dz)^{-1}~H_0~,$  
and with $k=0$ 
and $H(z) = H_0 E(z)$, 
\begin{equation}
E(z) = (dy/dz)^{-1}~
\end{equation}
(e.g. Weinberg 1972; Peebles 1993). Recent  
CMB measurements indicate that our universe 
has zero space curvature, $k=0$  
(e.g., Bennett et al. 2003, Spergel et al. 2003).    
Thus, in principle,
the data $y(z)$ can be used to empirically determine
the the $dy/dz$ and the dimensionless expansion rate $E(z)$.
This, in turn,
is related to cosmological parameters, such as dark energy,
and their redshift evolution as discussed in \S 6.  
For example, in a universe with quintessence (Caldwell, 
Dave, \& Steinhardt 1998), which has a time-independent
equation of state $w=P/\rho$,
$E^2(z) = \sum \Omega_i(1+z)^{n_i}$, 
where
$w_i=P_i/\rho_i$, and
$n_i =3(1+w_i)$ (see, for example, Turner \& White 1997;
Peebles \& Ratra 2003; or Daly \& Guerra 2002).  The  
deceleration parameter at the present epoch is 
$q_o= - (\ddot{a} a/ \dot{a})_o = 0.5 \sum \Omega_i(1+3w_i)$, when
$w_i$ is time independent.

A direct, empirical determination of the 
acceleration of the universe as a function of redshift can be
obtained from $y(z)$, without making any assumptions
about the nature or evolution of the ``dark energy.''
This can be done using the equation (Daly 2002)
\begin{equation}
-q(z) \equiv \ddot{a} a/\dot{a}^2 = 1~ +~ (1+z) ~(dy/dz)^{-1} (d^2y/dz^2)
\end{equation}
valid for $k=0$; if $k \neq 0$, another term 
$[kr(1+z)/(1-kr^2)](dr/dz)$ must be 
added to the right hand side of equation (6).

Equation (6) depends only upon the Robertson-Walker line element and
the relation  $(1+z) = a_o/a(t)$. Thus, this expression for 
$q(z)$ is valid for any homogeneous, isotropic expanding
universe in which $(1+z) = a_o/a(t)$, and is consequently quite
general and can be compared with any model to account for the
acceleration of the universe, as long as the model describes a
homogeneous isotropic expanding universe with the standard
relation between $z$ and $a(t)$.  The Robertson-Walker line element
is given by equation (4).  
A light ray emitted by a galaxy
traveling to us along the radial coordinate $r$
has $d \tau = d\theta=d\phi=0$.  The increment is along the
negative direction of $dr$  
so eq. (4) with k=0 implies that 
$a_o~dr = - (1+z)~dt$, or   $(dz/dt) = -a_o^{-1}(1+z)~(dr/dz)^{-1}$.  
Differentiating $(1+z) = a_o/a(t)$ with respect to time   
implies that $\dot{a} = -a_o~(1+z)^{-2} ~(dz/dt)$.
Substituting in for $(dz/dt)$, we find
$\dot{a} = (1+z)^{-1} (dr/dz)^{-1}$.  Differentiating
again with respect to time, we find
\begin{equation}
\ddot{a}=-(1+z)^{-2}~(dz/dt)~(dr/dz)^{-1}~[1+~(1+z)(dr/dz)^{-1}(d^2r/dz^2)]
~,
\end{equation}
which simplifies to eq. (6) using the expressions given here,
and the relation $y(z) = H_0 (a_or)$.  
For $k \neq 0$, equation (4) implies that 
$a_o~dr = - (1+z)\sqrt{(1-kr^2)}~dt$, or   
$(dz/dt) = -a_o^{-1}(1+z)\sqrt{(1-kr^2)}~(dr/dz)^{-1}$.  
Differentiating $(1+z) = a_o/a(t)$ with respect to time   
implies that $\dot{a} = -a_o~(1+z)^{-2} ~(dz/dt)$.
Substituting in for $(dz/dt)$, we find
$\dot{a} = (1+z)^{-1}\sqrt{(1-kr^2)} (dr/dz)^{-1}$.  Differentiating
again with respect to time, we find
$\ddot{a}=(a_o)^{-1}(1+z)^{-1}~(1-kr^2)~(dr/dz)^{-2}~
[1+(1+z)(dr/dz)^{-1}(d^2r/dz^2)] +(a_o)^{-1}~kr~(dr/dz)^{-1},$
thus $-q(z) = \ddot{a} a / \dot{a}^2 = [1+~(1+z)(dr/dz)^{-1}(d^2r/dz^2)]
+kr~(1+z) (1-kr^2)^{-1}~(dr/dz)$.

Equation (6) can in principle be
used to empirically determine the 
redshift at which the universe transitions from acceleration
to deceleration without requiring assumptions regarding
the nature and redshift evolution of the ''dark energy.''
The supernova and radio galaxy data allow a determination
of the dimensionless coordinate distance $y$ to each source,
at redshift $z$.  These data can then be used to determine
$dy/dz$, and $d^2y/dz^2$; these can then be substituted into
eq. (6) to determine q(z).

Since eqs. (5) and (6) are obtained without any assumptions regarding the
mass-energy components of the universe 
or their redshift evolution, they can 
be used to directly determine the dimensionless acceleration rate
$E(z)= (dy/dz)^{-1}$, which contains important information 
on the ``dark energy'' and its redshift evolution, and 
to determine the dimensionless 
acceleration parameter $q(z)$ directly from measurements
of $y(z)$.

In the determinations of $y(z)$ a value of ${\cal M_B}$ must be
adopted for the supernovae (see eq. 3), and a value of 
$\kappa$ must be adopted for radio galaxies (see eq. 1).  There are not
determined as a normalization using only low-redshift sources.
They are determined by fitting all of the data and solving for
the best fit values of these parameters.  Fits to the supernovae
data, the radio galaxy data, and the joint data set were run
for a variety of cases (see Table 2), such as a universe with 
quintessence (Q), or a rolling scalar field (SF).  
There are enough
supernovae that the value of ${\cal M_B}$ changes very little for different
fits to the supernovae data, and they change very little if the supernovae
data are considered separately or in conjunction with the radio galaxy
data.  Thus, values of $y$ for supernovae do not change with the
data set or model considered.  
The radio galaxy data best fit parameters for
$\kappa$ and $\beta$ 
change slightly depending upon whether
just the radio galaxies are considered, or whether the full data set
of radio galaxies plus supernovae are included.  Values of $y_s$ obtained
for the best fit value of $\kappa$ using radio galaxies 
alone in a universe with quintessence are listed, as well as the
values $y_j$ obtained using the best fit values of $\kappa$ and
$\beta$ for fits to the full data set of radio galaxies and
supernovae.  These values 
are listed in Table 1 and are considered and compared in 
the analyses of $E(z)$ and $q(z)$.

\section{The Numerical Differentiation Technique}

The key problem in this approach, of course, is that it requires a
numerical differentiation of typically noisy data, which is a cardinal
sin for any empirical scientist.  This, after all, is the reason why
all standard cosmological tests (e.g., the Hubble diagram) consist of
integrating the model equations to compare them with the measurements.
An additional problem is posed by the sparse and/or uneven coverage 
of the redshift range(s) of interest.
While a numerical differentiation of noisy data is in general not advisable,
it is certainly possible, and if done properly (in a statistical sense),
it can produce meaningful results within the limits of the available data.

Most numerical differentiation techniques explicitely or implicitely assume
that the data can be locally represented by some smooth (differentiable)
function, whose derivative is then defined analytically.  Typically this
local approximation is a low-order polynomial.  Thus, estimation of 
derivatives is coupled with the estimation of the function representing the
data themselves, in a self-consistent way.  Measurement errors can then
be propagated in the standard manner, leading to estimated uncertainties of
the fitted function values as well as the derivatives.  In our case, the
function to be approximated, along with its first and second derivatives,
is the dimensionless coordinate distance as a function of redshift, $y(z)$.
The situation is simplified by the fact that the errors in $z$ are negligible
in comparison to the errors in $y$, and thus the ordinary least-squares
approach can be used.

There are three sources of errors when evaluating any function fits to
noisy, finite data sets.  First, the errors of the individual data points:
the least-squares approach deals with them in a statistically optimal
fasion, provided that the quoted error bars are truly representative,
and that the deviations from the ``true'' underlying trend are drawn from
a normal distribution.  Second, if the fitted function is not
a good approximation to the true trend, the results may be systematically
biased.  Locally, any function can be approximated as a polynomial (or as
a Taylor series), and this becomes an issue of a having a sufficiently
high fit order to account for the shape (the curvature) of the observed trend
in the fitting interval.  Finally, in any finite data set there will be
some sample variance, i.e., a different draw of the same number of
measurements from the same underlying trend, with the same errors, will
produce slightly different results.  The effects of the sample variance
are minimised by having larger number of data points, and can be
estimated numerically for any given sample.

We choose a simple powers-of-$z$ polynomial approach, in order to be maximally
model-independent.  In principle, other basis functions could be used, 
but we do not see any advantages of such an approach in a situation where
the fits would be dominated by the noise and sparse sampling of the data.
We always fit to $y(z)$, and then derive the first and second derivatives
from the fit coefficients, and the local values of $E(z)$ and $q(z)$
using eqs. (5) and (6).  Uncertainties of the fit coefficients are then
propagated to derive the uncertainties in the fit values of $y(z)$, 
$E(z)$ and $q(z)$.  The fit values are always evaluated on a redshift
grid equally and densely spaced in either $z$ or $\log z$; this is just
a matter of convenience, as the values and the quality of the fits are not
affected.

A conceptually simplest approach would be to fit a polynomial to the entire
data set.  Unfortunately, low-order polynomials lack the flexibility
to represent the actual shapes of underlying cosmological models, leading
to seriously biased values of $E(z)$ and $q(z)$.  The fits are (by design)
optimised to fit the function $(y)$, and its derivatives are not
constrained directly.  Using higher order
polynomials helps in recovering the mean shapes of these functions,
but at the expense of greatly increased uncertainties, and typically
with some oscillatory behavior, characteristic of high-order polynomial fits.
For example, the $q(z)$ is generally a non-linear function of $z$,
so the fits of an order $> 3$ are needed; but in some cases, e.g.,
$\Omega_m = 1$ and $\Omega_\Lambda = 0$ cosmology, $q(z) = const.$,
which higher order polynomials cannot reproduce very easily in shape,
regardless of the increased errors for high-order terms.

A better method, which we adopted, is to fit the values of $y(z)$ locally,
in some limited redshift window of $\pm \Delta z$; within that interval,
data points are fitted with the weights inversely proportional to the
squares of their error bars.  In addition, at each end of the fitting
window we attach a Gaussian tapered region with a $\sigma(z) = 0.02$,
extending out to $2 \sigma$; the enclosed data points in the
tapered region have the weights lowered by the value of the Gaussian
wing at that point.  The purpose of this taper is to avoid fluctuations
caused by individual data points entering and leaving the fitting 
window, at the expense of a slight increase in the resulting fit
uncertainties (since the tapered points effectively get larger error
bars).  We established that the overall properties of the fits did
not change.  Finally, we require that there are at least 10 data points
in each fit, and increment the window slightly if necessary.

The tradeoff in this technique is that larger fitting windows lead to more
robust fits, at the expense of resolution in redshift and the introduction
of the same problems which plague the global polynomial fits, as described
above; while smaller fitting windows produce noisier fits because of a
smaller number of enclosed data points.  After some experimentation, we
concluded that windows with $\Delta z \approx 0.4$ seem to offer the optimal
compromise, but we also perform fits with other window sizes.

After some experimentation, we decided to fit second order polynomials
in each fitting window, as the minimal-assumption functions with defined
second derivatives (needed to evaluate the $q(z)$), which can also
accomodate any curvature in the data.  We verified that using linear
fits to obtain $y^\prime(z)$ and thus $E(z)$ does not produce improved 
results, and that increasing the local fit order to 3 increases the
formal errors without any significant benefits in terms of the fit
quality and accuracy.

Specific details of the fitting procedure are as follows.
Let the input data be $(z_i, y_i, \Delta y_i)$.  The fitting weights
are computed in a standard fashion as $1/\Delta y_i ^2$.
The fits are evaluated on an output reshift grid $z_j$, typically ranging
from 0.01 to 1.7, with a spacing of 0.01 or 0.005.  This is simply a choice
of convenience, since the fits can be evaluated anywhere in the redshift
regime covered by the data, and there is no reason to do it, say, just at
the values of the input $z_i$.  We note that since our fitting windows
$\Delta z$ are generally much larger than the output grid spacing, the
adjacent output fit values are $not$ independent.
For each output point, $z_j = z_0$, we select the input data in the
corresponding fitting window as described above, ranging from some
$i = i_{min}$ to some $i = i_{max}$; the fitting weights for points in
the Gaussian taper regions are adjusted appropriately.
We perform the ``centered'' fits by introducing the independent variable
$x_i = z_i - z_0$, and fit second order polynomials to $y(x)$ in the
range from $i_{min}$ to $i_{max}$.
We use the routine $fit$ from {\sl Numerical Recipes} (Press et al. 1992).
The output are the fitting coefficients $A, B, C$, where
$y(x) = A + B x + C x^2$.
Then at the centered fit value of $x = 0$, the best fit value for
$y(z) = A$,
the best fit value of its first derivative is
$dy/dz \equiv dy/dx = B$,
and the best fit value of the second derivative is 
$d^2y/dz^2 \equiv d^2y/dx^2 = 2 C$.
Thus, from eq. (5), $E(z) = 1/B$, 
and from eq. (6), $q(z) = -1 - (1+z) \times 2 C / B$.
The routine $fit$ also returns the covariance matrix, whose diagonal
elements give the uncertainties of the coefficients $A, B, C$, and 
their uncertainties are easily propagated to the fitting uncertainties
of $E(z)$ and $q(z)$.

We emphasize that our goal is not to evaluate a number of independent
measurements of $E(z)$ and $q(z)$ in the redshift range of interest
(the data in hand are not sufficient to do that for an interesting
number of points), but instead to outline the global trends presented
by the data.  Thus, we use relatively broad fitting windows with our
sliding window fit methodology.  The price we pay is the strong correlation
of fitted values on our densely spaced output grid, and these should
not be taken as independent measurements, but really as outlines of
the global trends.  In our model-independent approach there is an
implicit (and reasonable) assumption that the function $y(z)$ is smoothly
changing, with only a modest local curvature.  This implies that there
is a useful ``nonlocal'' information present in the data, which is
captured by our extended fitting windows, at a price of having only
2 or 3 ``independent'' measurements across our full redshift range.
With a richer input data set we could make the sliding windows smaller,
and increase the number of ``independent'' measurements of $E(z)$ and
$q(z)$ (i.e., our redshift resolution of the measured trends).

An alternative approach would be to bin the data in a modest number of
redshift bins, and perform an independent fit in each bin.  An
advantage of such a technique would be a set of truly independent
measurements of $E(z)$ and $q(z)$, but at a price of an increased
noise, since the information present by points outside each bin boundary
would be lost.  This would also lead to implied discontinuities in the
fit values of $y(z)$ at the bin edges, leading to implied infinite
derivatives, which is obviously unphysical.  Thus, we opt for a sliding
window fit approach, which is essentially a flexible and robust smoothing
technique, with a clear caveat that the output fit values are not
independent within the used $\Delta z$ range.

In exploring the numerical fitting and differentiation methodology, we
used artificial data sets with known, built-in cosmologies, in order to
evaluate the accuracy of the derived fits for $E(z)$ and $q(z)$.  
For most part, we generated artificial data sets mimicking what is
expected from SN measurements by the SNAP satellite 
(see, e.g., Aldering et al. 2003, or http://snap.lbl.gov/), namely a
set of 2000 measurements in the redshift interval from 0.1 to 1.7, with
combined (measurement + intrinsic) scatter of 7\% in dimensionless
coordinate distances.  The redshift distribution function was taken to
be proportional to the volume element divided by the redshift, which
roughly represents a combination of the expected SN rate history and
the SNAP selection function.  For most tests, we assumed the standard
Friedmann-Lemaitre cosmology with $\Omega_m = 0.3$ and $\Omega_\Lambda = 0.7$.
For some of the tests we changed some of these assumptions (the number
of the data points, the relative errors, or the underlying cosmology).

The overall process is illustrated in Fig. 3, on an example of our
pseudo-SNAP data set.  The global trends for $y(z)$, $E(z)$, and $q(z)$
are reproduced well, with the bias (systematic offsets) comparable to
the fit errors. The errors increase going towards the higher derivatives
and near the edges of the redshift intervals, as may be expected.

The effects of different fitting windows on the derived values $y(z)$
and $E(z)$ are illustrated in Fig. 4.
As expected, smaller values of $\Delta z$ lead to noisier fits, but the
overall trends agree within the errors.  For the present RG+SN data set,
we use $\Delta z$ of 0.4 or 0.6 in what follows.

Finally, we address the issue of the sample variance.  For our pseudo-SNAP
data, we simply generate a number of different realizations of the data
set, using the same assumptions.  In order to estimate the effects for
our RG+SN data set, we do the following.  We assume an underlying
cosmology, viz., $\Omega_m = 0.3$ and $\Omega_\Lambda = 0.7$.  Then,
we replace each $y(z)$ measurement with the value for this cosmology,
perturbed by a random amount drawn from a Gaussian distribution with
the $\sigma(y)$ equal to the quoted error bar.  We generate a number
of such pseudo-RGSN data sets, and perform the fits on each.  
The results are shown in Fig. 5.
For the RG+SN data set, the sample variance effects are comparable to
or smaller than the measurement errors for $y(z)$; and comparable to
the fitting uncertainties (which derive from the random errors of the
data) for $E(z)$.  As expected, the sample variance effects for the
pseudo-SNAP data set, which has many more data points, are effectively
negligible.

Estimates of the sample variance errors are made only rarely in the
published literature.  Our tests suggest that for many real-life data
sets in cosmology, these errors can be easily comparable to the fitting
uncertainties which derive from the random measurement errors, and thus 
in many cases the quoted confidence intervals may be underestimating the
total uncertainties.

\section{The Initial Results for E(z) and q(z)}

Using the procedure described above, the function $E(z)$ was obtained for
the full RG+SN data set using eq. (5), and is shown in Fig. 6. 
We used the values of
$y_j$ for RGs listed in Table 1, and the values for $y$ for SNe listed in 
Tables 3 and 4; a total of 78 SNe were used including the average values of
$y$ for the 14 SNe listed in Table 4, and values of $y$ for the remaining 64
SNe listed in Table 3.
The results are remarkably close to the 
currently popular ``concordance'' Friedmann-Lemaitre model
with $\Omega_{\Lambda} = 0.7$ and $\Omega_m = 0.3$.
We note, however, that in our analysis we did not assume that the universe
is described by a Friedmann-Lemaitre model at all.

As an internal consistency test, we computed the fits for the RG and SN samples
separately, using the values $y_s$ for the radio galaxies.   
The results are shown in Fig. 7. 
It is
clear that the values of $y$ obtained for radio galaxies
and supernovae agree very well for the redshifts where the data sets overlap.
And, in the redshift range where
the two samples overlap, the independent 
determinations of $E(z)$ agree to within 1-$\sigma$
(joint errors) or better.  It is notable that the RG data alone are consistent
with a constant $E(z) \approx 1.1$ for $z \sim 0.4 - 1.8$, although
the error bars are large.  If this trend remains as the error bars
decrease with better and more extensive data sets, it could be 
indicative of an actual cosmological trend, or (perhaps more likely) some
evolutionary effect or bias in the data.  This can not be sorted out
until more data are avialable.

Finally, we show in Fig. 8 what is probably the first direct estimate of
$q(z)$, obtained using eq. (6).  
The data, folded through our analysis procedure, are fully
consistent with the ``concordance'' model
with $\Omega_{\Lambda} = 0.7$ and $\Omega_m = 0.3$,
and suggest that indeed the universe transitions from decceleration to
acceleration at $z_T \gsim 0.3$, with a best fit value of 
$z_T \approx 0.45$.  
Again, we note that no assumptions about
the cosmological model have been made in deriving this trend.
This is a preliminary result, and we are clearly limited by the available
data at this time.  Our purpose here is mainly to illustrate the method,
but even so, the results are very encouraging.

\section{Implications of E(z) and q(z) for the Properties of the Dark Energy}

The acceleration parameter is $q(z) = -\ddot{a}a/\dot{a}^2 =
-(\ddot{a}/a)(\dot{a}/a)^{-2}$.  
The acceleration of the universe is described by    
\begin{equation}
(\ddot{a}/a) = -{4 \pi G \over 3}~ \sum (\rho_i + 3 p_i)
= -{4 \pi G \over 3}~ \sum \rho_i(1 + 3 w_i)
\label{acceq}
\end{equation} where $p_i$ is the pressure, $\rho_i$ is the 
mean mass-energy density, $w_i$ is the equation
of state of $i\mbox{th}$ component, $w_i = p_i /\rho_i$.  
In addition, for $k=0$, 
\begin{equation}
(\dot{a}/a)^2 = {8 \pi G \over 3}~ \sum \rho_i~.
\label{eqadot}
\end{equation}

Thus, $q(z) = 0.5 \sum \rho_i(1+3w_i)/ \sum \rho_i$ when 
$k=0$, and $E^2(z)=\sum \rho_i/\rho_{co}$, where $\rho_{co}$
is the critical density at the present epoch, 
$\rho_{co} \equiv (3/8 \pi G)H_o^2$ = $\rho_{mo} +\rho_{Eo}$,
and the present epoch mean mass-energy density of
non-relativistic matter and dark energy are 
$\rho_{mo}$ and $\rho_{Eo}$ respectively.  Non-relativistic
matter evolves with redshift as $(1+z)^3$.  Let the dark
energy evolve with redshift as $f_E(z)$, then, it is easy
to show that $E^2(z) = \Omega_m(1+z)^3 + (1-\Omega_m)f_E(z)$,
where $\Omega_m = \rho_{mo}/\rho_{co}$, and $\Omega_E = 
\rho_{Eo}/\rho_{co} = 1-\Omega_m$. 
Hence, if the current contribution of non-relativistic matter
$\Omega_m$ can be determined, then $E(z)$ can be used 
to determine the redshift evolution of the dark energy $f_E(z)$.  
Similarly, $q(z) = 0.5[\Omega_m+(1-\Omega_m)(1+3w)(1+z)^{-3}~f_E(z)]
/[\Omega_m+(1-\Omega_m)(1+z)^{-3}f_E(z)]$.

Now, for quintessence, 
$\rho_i = \rho_{i,o}(1+z)^n_i$ when $w_i$ is
constant, where $n_i=3(1+w_i)$.   This follows from the 
mass-energy conservation of each component, which implies 
\begin{equation}
\dot{\rho_i} = -3(\rho_i + p_i)(\dot{a}/a)
\label{conseq}
\end{equation}
(e.g. Peebles 1993).
When the equation of state $w_i$ does not change with time,
the solution to this equation is 
$\rho_i = \rho_{i,o}(1+z)^{3(1+w_i)}$, where $(1+z)=a_o/a$.  
Thus, a component with equation of state $w_i$ and
present mean mass-energy density $\rho_o$ will have 
a mean mass-energy density at redshift $z$ of
$\rho=\rho_o(1+z)^{n_i}$, where $n_i=3(1+w_i)$.

With two important components at low redshift, a non-relativistic
component $\rho_{m}$ that includes baryons and the dark matter in galaxies
and clusters of galaxies, and a dark energy component $\rho_E$, we
have $\rho_m = \rho_{mo}(1+z)^3$ and $\rho_E = \rho_{Eo}
(1+z)^n$.  Now, at zero redshift the total
density is equal to the critical density $\rho_{co}=\rho_{mo}
+\rho_{Eo}$ and since $\rho_{mo}/\rho_{oc} \equiv \Omega_m$, 
the acceleration parameter may be written
\begin{equation}
q(z) = 0.5
\left( {\Omega_m +(1-\Omega_m)(1+3w)(1+z)^{3w} \over
\Omega_m +(1-\Omega_m)(1+z)^{3w}}\right)~.
\end{equation}
The universe is decelerating when the sign of $q(z)$ is positive.
The sign of the denominator is always positive, and the 
numerator may be either positive or negative depending upon
the value of $w$.  The universe will go
from a state of acceleration to a state of deceleration if
the dark energy has properties like that of quintessence when 
the numerator of equation (8) is positive, which occurs
a the transition redshift $z_T$ given by
\begin{equation}
z_T = \left[{-\Omega_m} \over {(1+3w)(1-\Omega_m)}\right]^{1/(3w)}-~1~.
\end{equation} 
This transition redshift is plotted as a function of the
equation of state $w$ in Figure 9.  Clearly, as the transition
redshift increases, the value of $\Omega_m$ must decrease, or
the equation of state $w$ exhibit redshift evolution.

Similarly, for quintessence, $E^2(z) = 
(1+z)^3[\Omega_m+(1-\Omega_m)(1+z)^{3w}]$.  Some
lines representing quintessence with $w=-1$ (i.e. 
a cosmological constant) are included
in the figures.

Our preliminary results give a limit to the transition redshift
$z_T \gsim 0.3$, with the best fit estimate $z_T \approx 0.45$.
Assuming $\Omega_m = 0.3$, these translate to $w \lsim -0.55$ and
$-2.3 \lsim w \lsim -0.65$ (see Figure 9).  
With better data sets in the future, we should be
able to improve on these limits.

Note that $q(z)$ is an important input into the Statefinder diagnostic
presented by Sahni et al. (2002) and Alam et al. (2003);  these authors
discuss one way in which $q(z)$ may be used to determine $w(z)$ and 
the Statefinder pair (r,s).

\section{Summary and Discussion}

We presented here a set of dimensionless coordinate distances for 20
RGs, spanning the redshift range 0.43 to 1.79 (with one source at
$z = 0.056$).  These measurements supplement and extend to the cosmologically
interesting redshift range the distances available
for SNe, which currently reach only to $z = 0.97$ (with one source at
$z = 1.70$).

The determination of the dimensionless coordinate distances to RGs
and SNe are completely independent, and are based on completely different
physics.  Yet, the two data sets agree very well in the overlap redshift
range, as shown here, and as shown by previous work
(e.g., Podariu et al. 2002;
Daly \& Guerra 2002; Guerra, Daly, \& Wan 2000). 
This is very encouraging: there is a great value in
being able to measure the same physical quantity (here the coordinate distances
as a function of redshift) using different and independent tracers and methods.
The general agreement we see between the SN and RG data sets suggests that
neither method is dominated by some substantial, as yet unknown systematic
errors.  Together, the two data sets can be used in cosmological tests
with a greater power than each data set separately.

The dimensionless coordinate distances $y(z)$ can be used to empirically
determine the dimensionless expansion and deceleration rates as functions
of redshift, $E(z)$ and $q(z)$, without assuming any particular cosmological
model.  While the traditional cosmological tests integrate the expressions
for these functions provided by the models (e.g., the standard
Friedmann-Lemaitre models) and determine the model parameters from the fits,
we develop a complementary procedure whereby these functions can be derived
directly from the data by differentiating the $y(z)$ trend.  We apply a
particular, simple numerical procedure to this task, and derive the trends
of both $E(z)$ and -- for the first time -- $q(z)$ directly from the data.

Our estimates of $E(z)$ are in an excellent agreement with those obtained
from other methods, e.g., the CMBR fluctuations,large-scale
structure,  high-$z$ SNe and
radio galaxies using traditional analyses, 
etc., even through they are obtained in a completely different
and independent manner.  In particular, the data are consistent with the
``concordance'' cosmology, i.e., Friedmann-Lemaitre models with 
$\Omega_m = 0.3$, and $\Omega_{\Lambda} = 0.7$.
While these results are clearly very preliminary, and meant primarily
to illustrate the method, the good agreement with other approaches is
very encouraging.

We are currently limited by the amount and quality of the available data for
both SNe and RGs.  Nevertheless, there are great
prospects for advances in precision cosmology, e.g., large sets of
high-quality measurements of SNe from the SNAP satellite
(e.g., Aldering et al. 2003),
or from large ground-based experiments such as the
ESSENCE (Stubbs 2002; see also http://www.ctio.noao.edu/wproject/)
or LSST in the future (Tyson et al. 2002, 2003).
Such data sets could certainly support differentiation of distance vs.
redshift trends, leading to considerably more robust direct determinations
of the expansion and acceleration rates as functions of redshift.
In addition, new radio galaxy data is being obtained.

As the observational situation improves, direct estimates 
of $E(z)$ and $q(z)$ can be used to understand the properties and redshift
evolution of different flavors of dark energy, determine the redshift at which
the universe transitions from acceleration to deceleration, and may help 
elucidate any systematic errors that might be lurking in the RG or SN methods
of constraining cosmological parameters.

\acknowledgments
It is a pleasure to thank Megan Donahue, Eddie
Guerra, Matt Mory, Chris O'Dea, Paddy Padmanabhan, Bharat Ratra, 
and Varun Sahni
for helpful comments and discussions, Saul Perlmutter for
sending electronic files of supernova data, 
and Adam Reiss for 
providing us with supernova data modified to be on the same
scale as that provided by Saul Perlmutter. 
This work 
was supported in part by the U. S. National Science Foundation
under grants AST-0096077 and AST-0206002, and Penn State University (RAD),
and by the Ajax Foundation (SGD).

\begin{deluxetable}{llllll}
\tablewidth{0pt}

\tablecaption{Radio Galaxy Dimensionless Coordinate Distances\label{RGyofz}}
\tablehead{
\colhead{Source} & \colhead{ Redshift } & \colhead{$y_j$} & \colhead{$\sigma(y_j)$} & \colhead{$y_s$} & \colhead{$\sigma(y_s)$}
\\}
\startdata   
3C405&0.056&0.056&0.010&0.057&0.011\\ 
3C244.1&0.430&0.445&0.071&0.462&0.079\\ 
3C330&0.549&0.400&0.066&0.431&0.076\\ 
3C427.1&0.572&0.319&0.051&0.319&0.054\\ 
3C337&0.630&0.600&0.071&0.630&0.080\\ 
3C55&0.720&0.606&0.071&0.680&0.085\\ 
3C247&0.749&0.625&0.069&0.660&0.077\\ 
3C265&0.811&0.667&0.081&0.731&0.093\\ 
3C325&0.860&0.818&0.149&0.885&0.162\\ 
3C289&0.967&0.681&0.108&0.722&0.122\\ 
3C268.1&0.974&0.780&0.127&0.855&0.149\\ 
3C280&0.996&0.703&0.111&0.758&0.128\\ 
3C356&1.079&0.842&0.151&0.979&0.188\\ 
3C267&1.144&0.753&0.126&0.837&0.150\\ 
3C194&1.190&1.141&0.205&1.251&0.239\\ 
3C324&1.210&0.996&0.251&1.081&0.291\\ 
3C437&1.480&0.849&0.206&0.992&0.260\\ 
3C68.2&1.575&1.477&0.386&1.717&0.484\\ 
3C322&1.681&1.167&0.249&1.356&0.316\\ 
3C239&1.790&1.246&0.257&1.419&0.318\\    
\enddata
\end{deluxetable}

\begin{deluxetable}{lllllll}
\tablewidth{0pt}

\tablecaption{Best Fit Parameters \label{BFP}}
\tablehead{
\colhead{Model} & \colhead{Fit To} & 
\colhead{${\cal M_B}$} & \colhead{$\kappa$} & \colhead{$\beta$}
& \colhead{${\chi^2 \over dof}$(SN)} & \colhead{${\chi^2 \over dof}$(RG)} \\}
         
\startdata

Q & SN/RG only (s) & $23.91 \pm 0.03$ & $8.88 \pm 0.05$ & $1.70 \pm 0.04$ &
56.2/50& 16.53/16 \\
Q & SN+RG (j)  & $23.95 \pm 0.03$ & $8.81 \pm 0.05$ & $1.75 \pm 0.04$ 
& 74.1/68 &74.1/68\\
SF & SN/RG only (s) & $23.94 \pm 0.03$ & $8.90 \pm 0.05$ & $1.70 \pm 0.04$ 
& 56.7/50 & 16.7/16\\
SF & SN+RG (j) & $23.95 \pm 0.03$ & $8.81 \pm 0.05$ & $1.80 \pm 0.03$ 
& 74.1/68 & 74.1/68\\

\enddata
\end{deluxetable}

\begin{deluxetable}{lllll}
\tablewidth{0pt}

\tablecaption{Supernovae Ia Dimensionless Coordinate Distances \label{SNyofz}}
\tablehead{
\colhead{Source} & 
\colhead{ Redshift } & \colhead{$y_s$} & \colhead{$\sigma(y_s)$} &\colhead{ref}\\}

\startdata

1996C&0.009&0.008&0.0012&R98\\ 
1995D&0.012&0.011&0.0011&R98\\ 
1992al&0.014&0.013&0.0013&P99\\ 
1992al&0.014&0.013&0.0009&R98\\ 
1995ak&0.016&0.015&0.0014&R98\\ 
1994S&0.016&0.015&0.0010&R98\\ 
1992bo&0.018&0.019&0.0016&R98\\ 
1992bc&0.020&0.017&0.0016&P99\\ 
1992bc&0.020&0.020&0.0012&R98\\ 
1995ac&0.022&0.023&0.0018&R98\\ 
1994M&0.024&0.026&0.0022&R98\\ 
1993H&0.025&0.024&0.0029&R98\\ 
1992ag&0.026&0.026&0.0028&R98\\ 
1992ag&0.026&0.029&0.0027&P99\\ 
1992P&0.026&0.026&0.0029&P99\\ 
1992P&0.026&0.030&0.0020&R98\\ 
1995bd&0.028&0.034&0.0033&R98\\ 
1999O&0.030&0.028&0.0026&P99\\ 
1992bg&0.035&0.038&0.0038&R98\\ 
1994T&0.036&0.035&0.0035&R98\\ 
1992bg&0.036&0.034&0.0033&P99\\ 
1992bl&0.043&0.043&0.0036&P99\\ 
1992bl&0.043&0.038&0.0026&R98\\ 
1992bh&0.045&0.051&0.0040&R98\\ 
1992bh&0.045&0.052&0.0046&P99\\ 
1995E&0.049&0.049&0.0031&R98\\ 
1990af&0.050&0.052&0.0044&P99\\ 
1993ag&0.050&0.054&0.0050&P99\\ 
1993ag&0.050&0.048&0.0038&R98\\ 
1990af&0.050&0.042&0.0031&R98\\ 
1993O&0.052&0.053&0.0030&R98\\ 
1993O&0.052&0.050&0.0042&P99\\ 
1992bs&0.063&0.068&0.0057&P99\\ 
1992bs&0.064&0.069&0.0045&R98\\ 
1993B&0.071&0.071&0.0065&P99\\ 
1992ae&0.075&0.074&0.0058&R98\\ 
1992ae&0.075&0.074&0.0068&P99\\ 
1992bp&0.079&0.068&0.0057&P99\\ 
1992bp&0.080&0.069&0.0044&R98\\ 
1992br&0.087&0.087&0.0077&R98\\ 
1992aq&0.101&0.096&0.0067&R98\\ 
1992aq&0.101&0.101&0.0107&P99\\ 
1996ab&0.124&0.124&0.0073&R98\\ 
1997I&0.172&0.151&0.0126&P99\\ 
1997N&0.180&0.169&0.0133&P99\\ 
1996J&0.300&0.319&0.0357&R98\\ 
1997ac&0.320&0.291&0.0243&P99\\ 
1994F&0.354&0.361&0.0549&P99\\ 
1994am&0.372&0.337&0.0312&P99\\ 
1994an&0.378&0.389&0.0663&P99\\ 
1996K&0.380&0.337&0.0310&R98\\ 
1995ba&0.388&0.399&0.0369&P99\\ 
1995aw&0.400&0.346&0.0304&P99\\ 
1997am&0.416&0.376&0.0349&P99\\ 
1994al&0.420&0.372&0.0430&P99\\ 
1994G&0.425&0.305&0.0690&P99\\ 
1996E&0.425&0.344&0.0449&R98\\ 
1997Q&0.430&0.373&0.0311&P99\\ 
1996U&0.430&0.497&0.0574&R98\\ 
1997ce&0.440&0.375&0.0304&R98\\ 
1995az&0.450&0.358&0.0380&P99\\ 
1996cm&0.450&0.485&0.0515&P99\\ 
1997ai&0.450&0.414&0.0574&P99\\ 
1995aq&0.453&0.484&0.0559&P99\\ 
1992bi&0.458&0.469&0.0994&P99\\ 
1995ar&0.465&0.516&0.0715&P99\\ 
1997P&0.472&0.464&0.0409&P99\\ 
1995K&0.478&0.459&0.0353&R98\\ 
1995ay&0.480&0.431&0.0478&P99\\ 
1996ci&0.495&0.402&0.0354&P99\\ 
1995as&0.498&0.602&0.0695&P99\\ 
1997cj&0.500&0.442&0.0340&R98\\ 
1997H&0.526&0.456&0.0423&P99\\ 
1997L&0.550&0.530&0.0613&P99\\ 
1996cf&0.570&0.469&0.0477&P99\\ 
1996I&0.570&0.499&0.0430&R98\\ 
1997af&0.579&0.513&0.0522&P99\\ 
1997F&0.580&0.508&0.0541&P99\\ 
1997aj&0.581&0.428&0.0436&P99\\ 
1997K&0.592&0.785&0.1340&P99\\ 
1997S&0.612&0.554&0.0538&P99\\ 
1995ax&0.615&0.439&0.0507&P99\\ 
1997J&0.619&0.580&0.0750&P99\\ 
1996H&0.621&0.535&0.0434&R98\\ 
1995at&0.655&0.445&0.0432&P99\\ 
1996ck&0.656&0.510&0.0660&P99\\ 
1997R&0.657&0.575&0.0611&P99\\ 
1997G&0.763&0.725&0.1772&P99\\ 
1996cl&0.828&0.760&0.1892&P99\\ 
1997ap&0.830&0.652&0.0664&P99\\ 
1997ck&0.970&0.844&0.1169&R98\\ 
1997ff&1.700&0.967&0.1517&R02\\ 
\enddata 
\end{deluxetable}

\begin{deluxetable}{llll}
\tablewidth{0pt}

\tablecaption{Average Values of y for Supernovae Listed by Both R98 
and P99 \label{SNyofz2}}
\tablehead{
\colhead{Source} & 
\colhead{ Redshift } & \colhead{$y_s$} & \colhead{$\sigma(y_s)$} \\}

\startdata
1990af&0.050&0.047&0.0053\\ 
1992ae&0.075&0.074&0.0090\\ 
1992ag&0.026&0.027&0.0039\\ 
1992al&0.014&0.013&0.0016\\ 
1992aq&0.101&0.098&0.0126\\ 
1992bc&0.020&0.019&0.0020\\ 
1992bg&0.036&0.036&0.0050\\ 
1992bh&0.045&0.051&0.0061\\ 
1992bl&0.043&0.040&0.0044\\ 
1992bp&0.079&0.069&0.0072\\ 
1992bs&0.063&0.069&0.0073\\ 
1992P&0.026&0.028&0.0035\\ 
1993ag&0.050&0.051&0.0063\\ 
1993O&0.052&0.052&0.0051\\ 
\enddata 
\end{deluxetable}

\begin{figure}
\includegraphics[width=150mm]{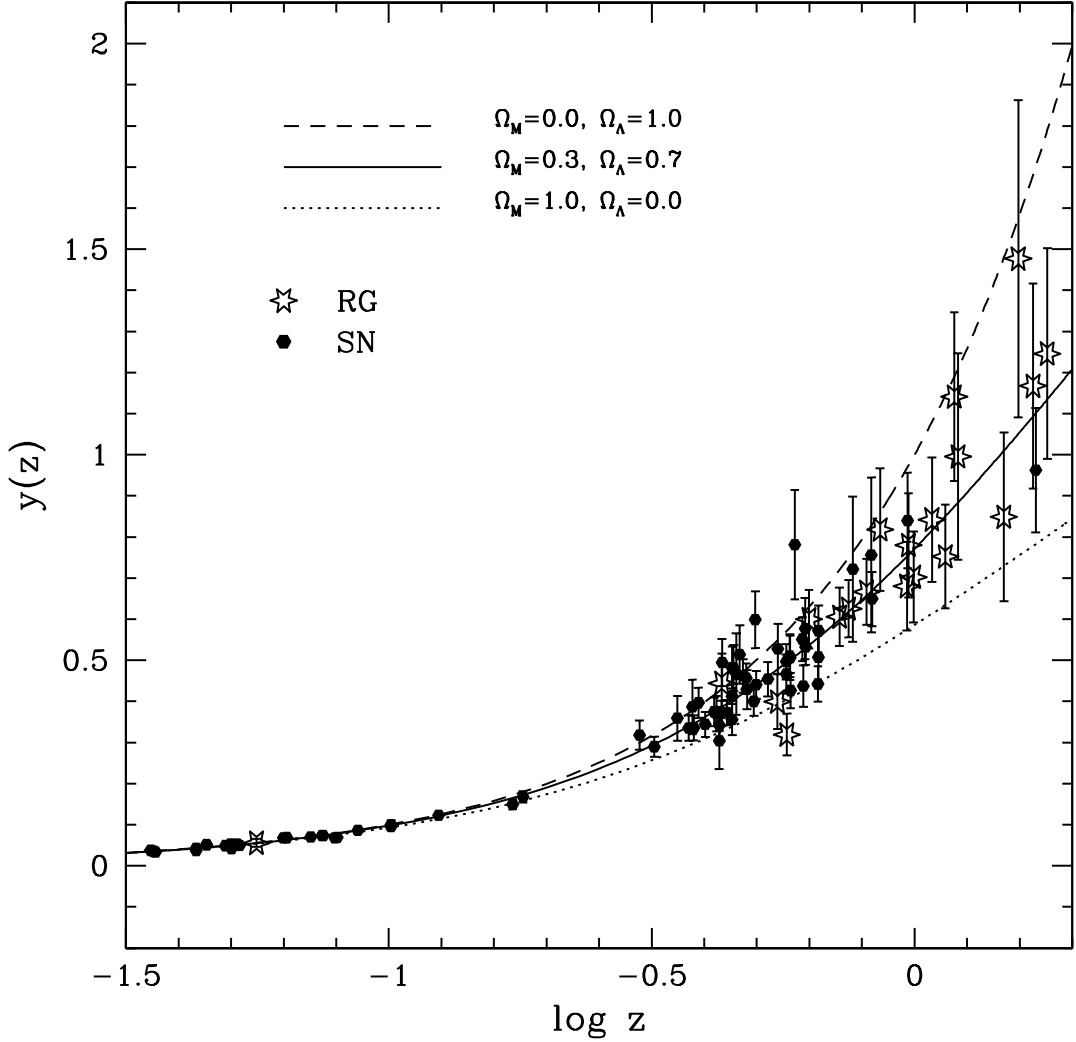}
\caption{Dimensionless coordinate distances $y(z)$ to 20 radio 
galaxies and 78 supernovae as a function 
of log z. Note that the determinations
of $y(z)$ have been made using best fit value of
${\cal M_B}$ obtained for the full data set of 78 supernovae and
20 radio galaxies, and the best fit value of $\kappa$ obtained
using the full data set ($y_j$ for radio galaxies).  
Radio galaxies are shown as open stars and
supernovae are shown as 
solid circles. Very similar results obtain when values of $y_s$ for
radio galaxies are shown, as in Fig. 2. }
\end{figure}
\clearpage

\begin{figure}
\includegraphics[width=150mm]{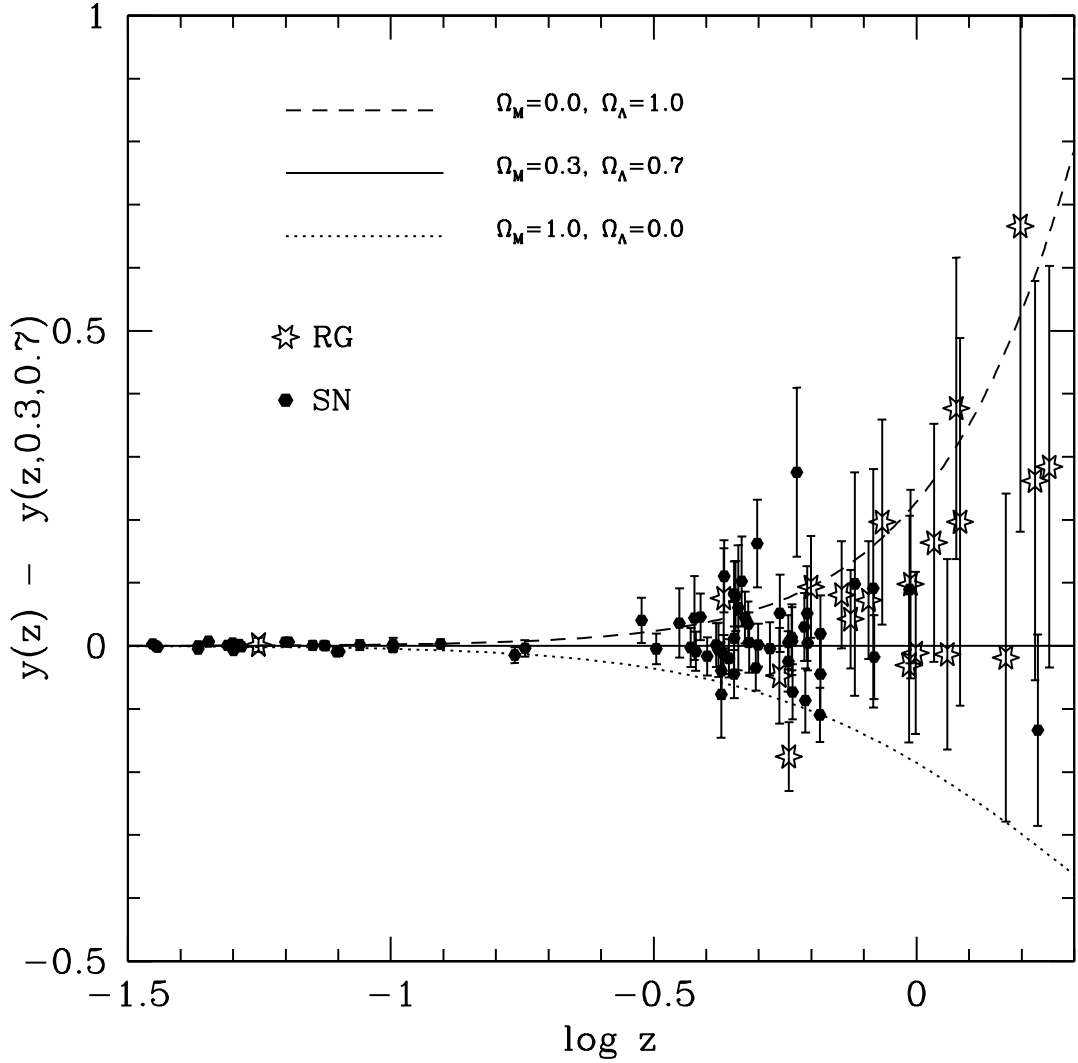}
\caption{The residuals between $y(z)$ and those expected in 
a universe with $\Omega_m = 0.3$
and $\Omega_{\Lambda} = 0.7$, where $y(z)$ is 
the dimensionless coordinate distance,
shown as a function of $\log z$.  Values of $y_s$,
obtained using the best fit values of $\kappa$ 
and $\beta$ determined using 
radio galaxies alone, are shown.  The results obtained
when best fit values to the full data set are used
($y_j$ for radio galaxies) are very 
similar, as shown in Fig. 1.  
Radio galaxies are shown as open stars and
supernovae are shown as 
solid circles. } 
\end{figure}
\clearpage

\begin{figure}
\includegraphics[width=90mm]{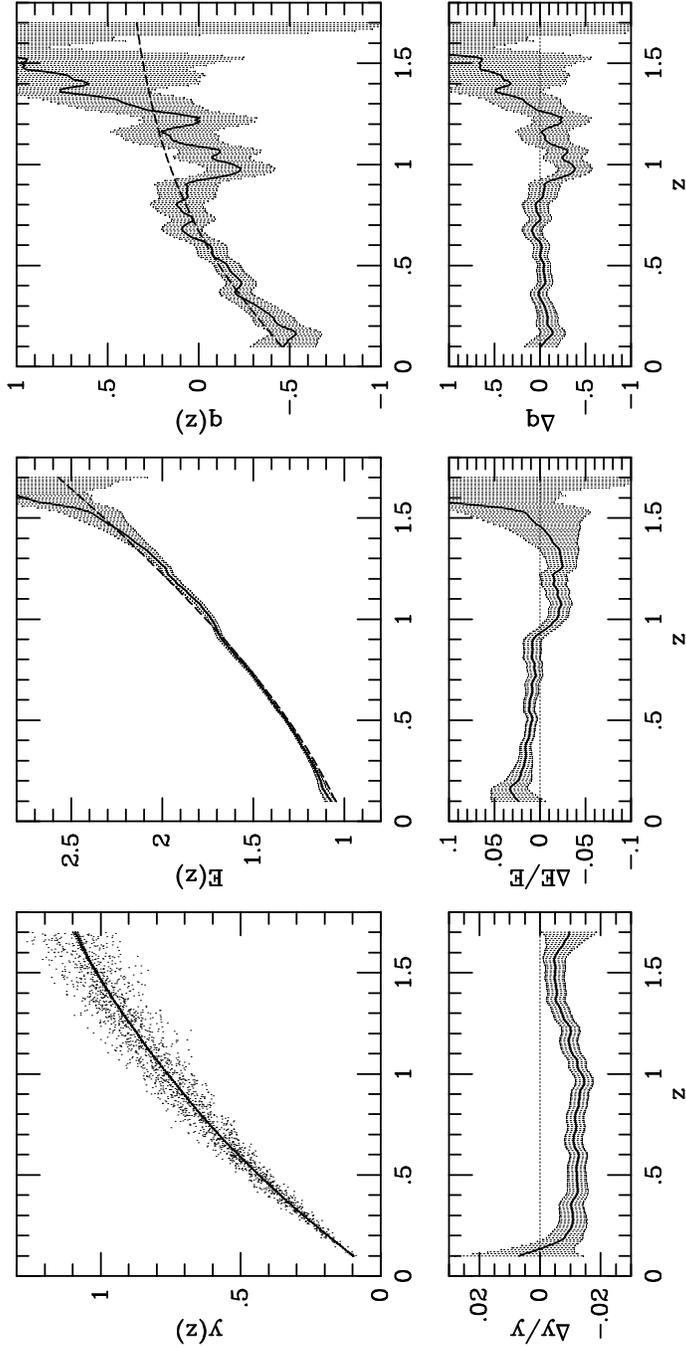}
\caption{
An example of the sliding-window fitting results for a pseudo-SNAP data
set with 2000 data points.  We assumed the cosmology with $\Omega_m = 0.3$
and $\Omega_{\Lambda} = 0.7$, and the relative errors $\Delta y / y$ drawn 
from a Gaussian distribution with $\sigma = 7$\%.  A  window function with
$\Delta z = 0.4$ was used.  
The top left panel shows the input data (dots) along with the recovered $y(z)$
trend.
The bottom left panel shows the $y(z)$ fit residuals from the values
corresponding to the assumed cosmology.
The middle panels show the fits for $E(z)$ (top) and its residuals (bottom),
and the right panels the equivalent for $q(z)$.
In all cases the thick line shows the fit values, and the hashed area indicates
the $\pm 1\sigma$ uncertainties.
In the top middle and right panels, the dashed lines show the theoretical
values for the assumed cosmology.
} 
\end{figure}
\clearpage

\begin{figure}
\includegraphics[width=150mm]{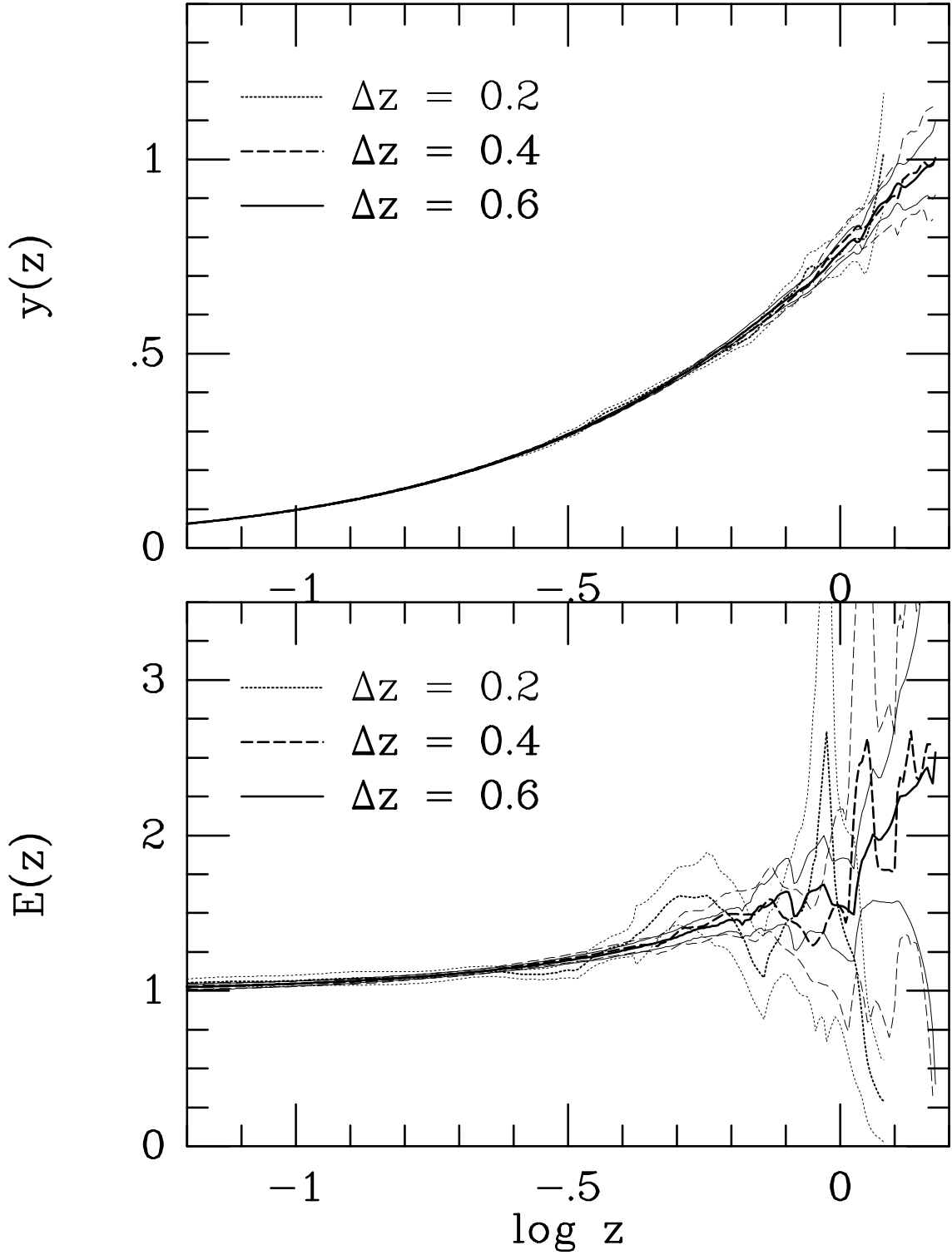}
\caption{
Comparison of results obtained for $E(z)$ when different fitting
window functions are applied to the full data set of 20 radio galaxies and
78 supernovae; values of $y_j$ for the radio galaxies were used for these fits.
Examples of three different window functions are shown with different line
types, as indicated.  The fit values are indicated by the thick lines, whereas
the corresponding thin lines indicate the $\pm 1\sigma$ range.
} 
\end{figure}
\clearpage

\begin{figure}
\includegraphics[width=100mm]{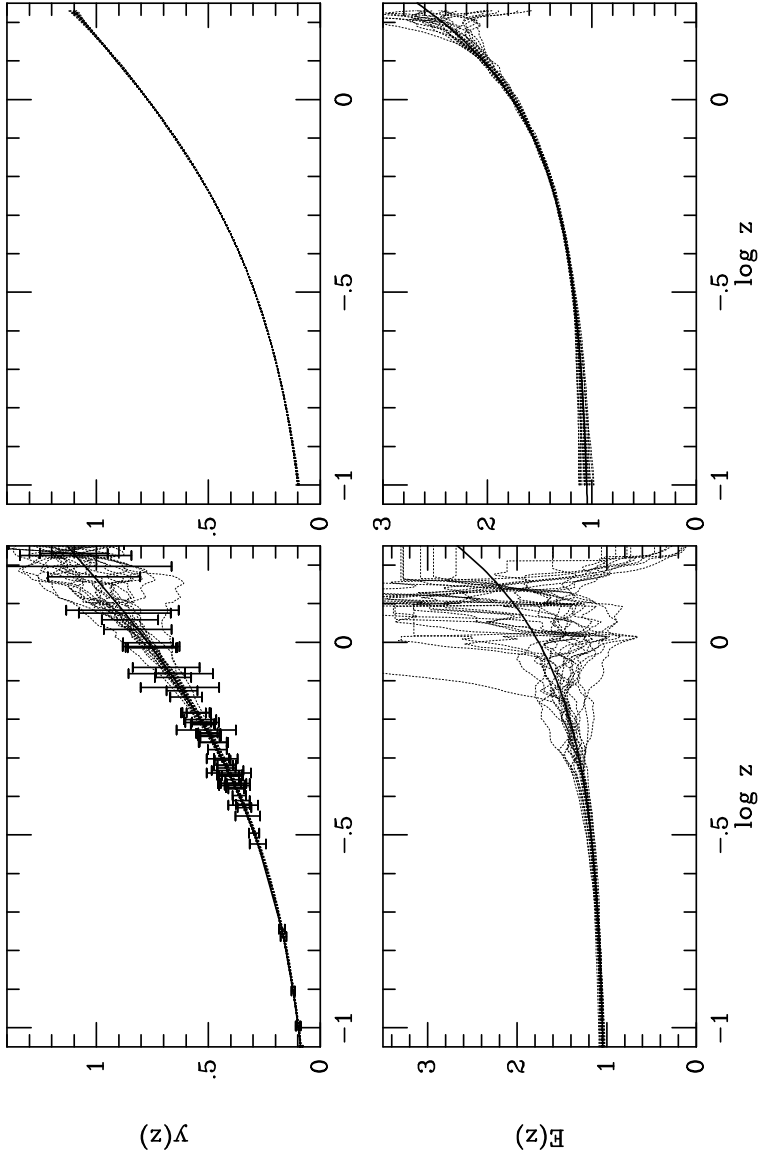}
\caption{
Modeling of the sample variance effects for the RG+SN sample (left) and
the pseudo-SNAP sample (right); values of $y_j$ were used for
the radio galaxies.  A cosmology with 
$\Omega_m = 0.3$, and $\Omega_{\Lambda} = 0.7$ was assumed, and is shown
with the solid lines in each panel.  The placement and the error bars of the
SN+RG data points are also indicated in the top left panel.  Each dotted line
represents a fit from a single random realization of the mock data sets, as
described in the text.  Their spread at a given redshift is indicative of the
sample variance errors.  These are obviously much more significant for the
smaller-$N$, RG+SN data set, than for the much larger pseudo-SNAP data set.
} 
\end{figure}
\clearpage

\begin{figure}
\includegraphics[width=150mm]{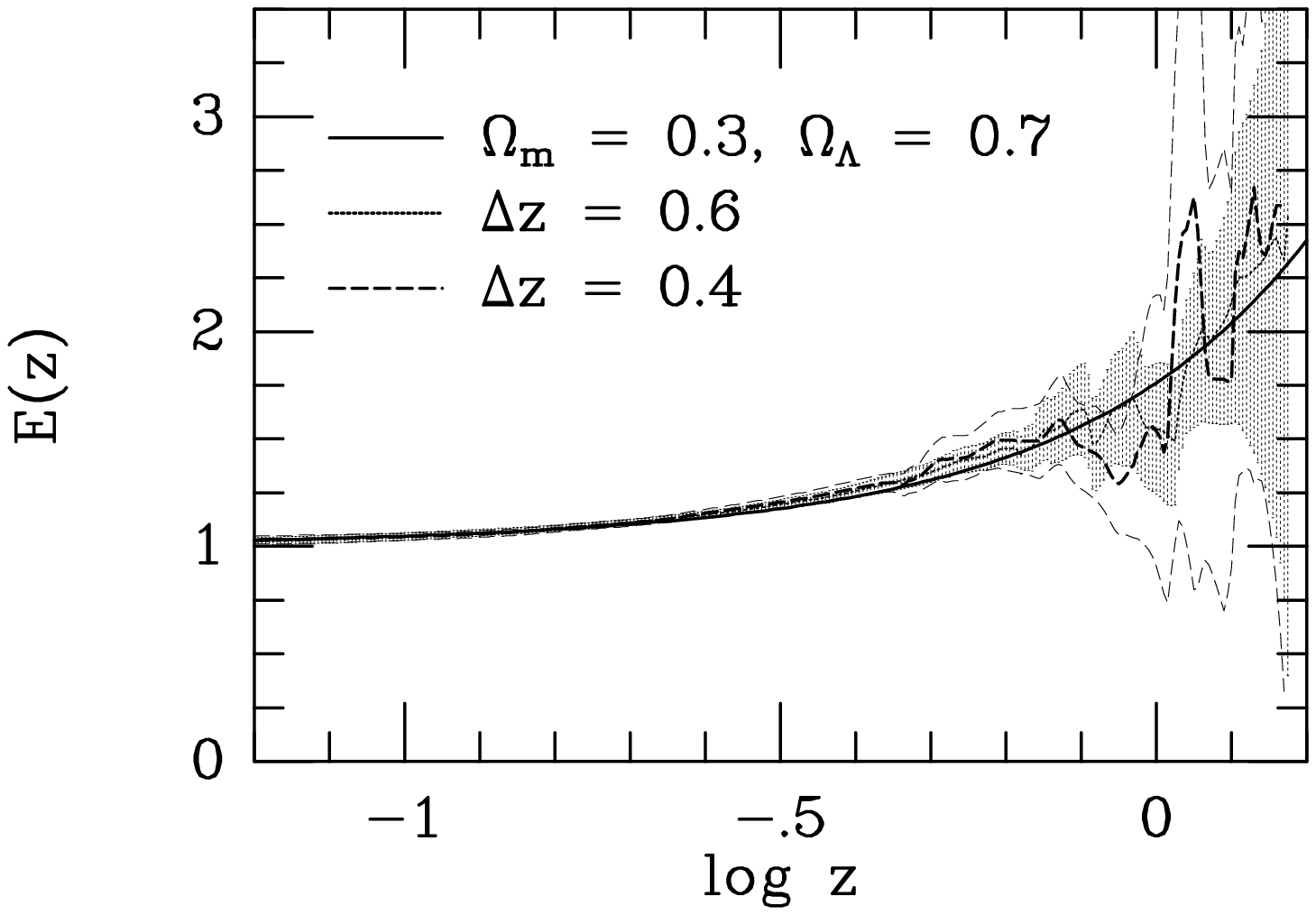}
\caption{
A first look at $E(z)$ for the full data set, with values of 
$y_j$ used for the radio galaxies.  
Fits with $\Delta z = 0.6$ are shown with the thick dotted line,
with the dotted hash indicating the $\pm 1\sigma$ range,
and fits with $\Delta z = 0.4$ are shown with the thick dashed line,
with the thin dashed lines indicating the $\pm 1\sigma$ range.  
The thick solid line shows the trend for the 
$\Omega_m = 0.3$, and $\Omega_{\Lambda} = 0.7$ Friedmann-Lemaitre cosmology. 
} 
\end{figure}
\clearpage

\begin{figure}
\includegraphics[width=150mm]{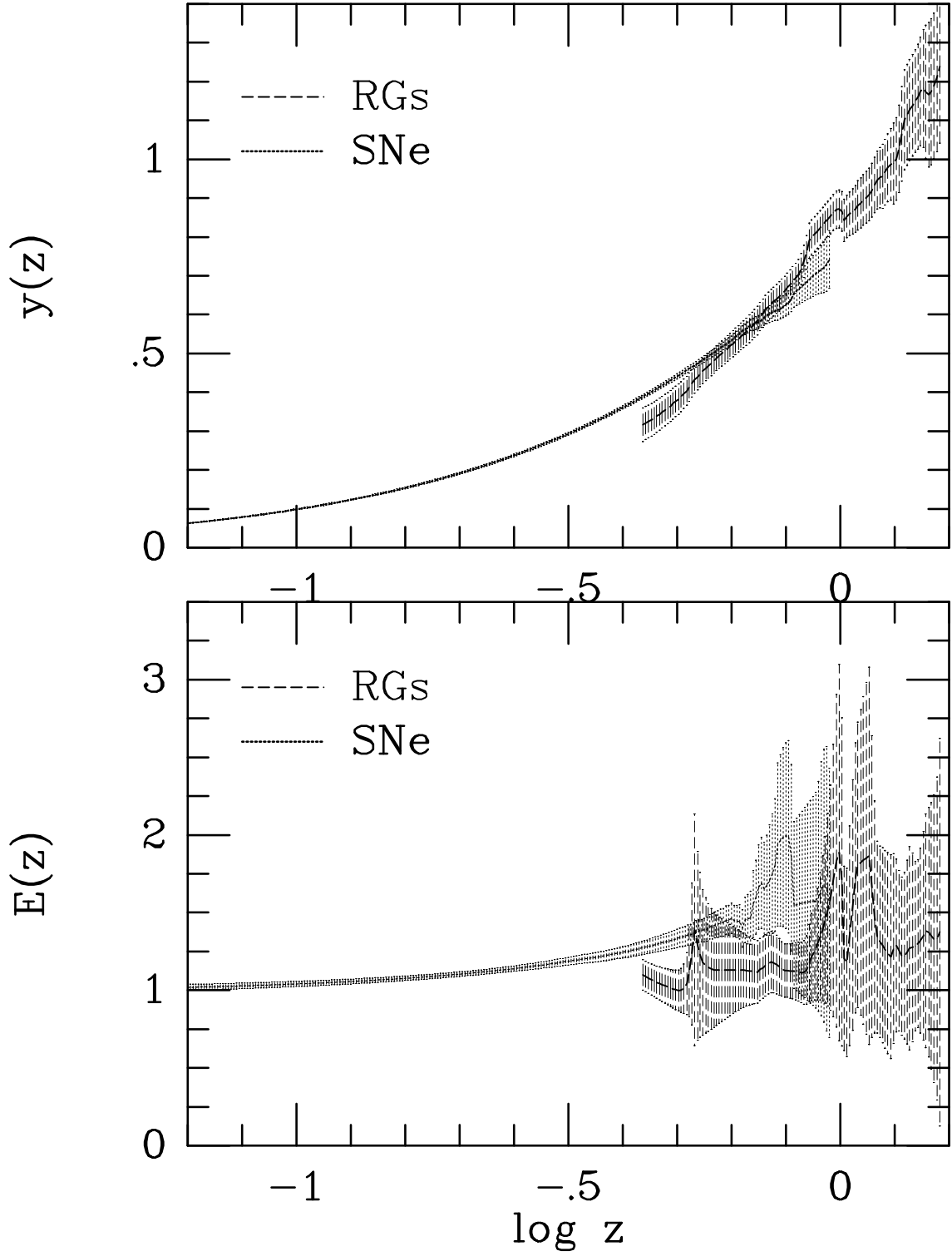}
\caption{A comparison of the fits for SN and RG data sets separately; values
of $y_s$ for RGs were used for these fits.  The SN data are shown as the dotted
line and the dotted hatched area (the best fit values and the $\pm 1\sigma$
range).  The corresponding fits for RGs are shown with the dashed line and
hatched area.
} 
\end{figure}
\clearpage

\begin{figure}
\includegraphics[width=150mm]{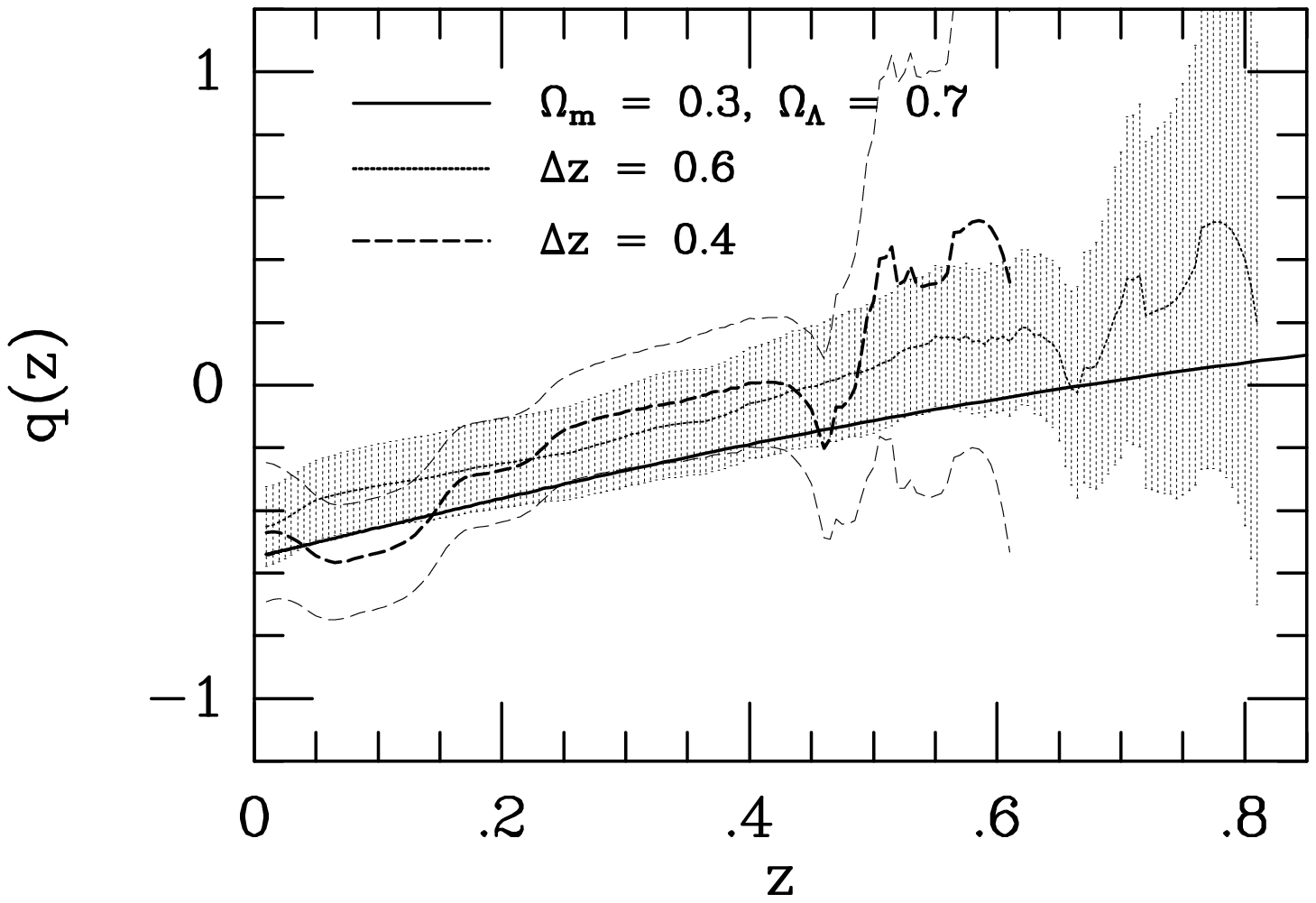}
\caption{
A first look at $q(z)$, obtained for the full data set, with
values of $y_j$ used for the radio galaxies.  
Fits with $\Delta z = 0.6$ are shown with the thick dotted line,
with the dotted hash indicating the $\pm 1\sigma$ range,
and fits with $\Delta z = 0.4$ are shown with the thick dashed line,
with the thin dashed lines indicating the $\pm 1\sigma$ range.  
The thick solid line shows the trend for the 
$\Omega_m = 0.3$, and $\Omega_{\Lambda} = 0.7$ Friedmann-Lemaitre cosmology. 
The fits become too noisy to be useful past about $z \sim 0.8$ in this
data set. 
} 
\end{figure}
\clearpage

\begin{figure}
\includegraphics[width=150mm]{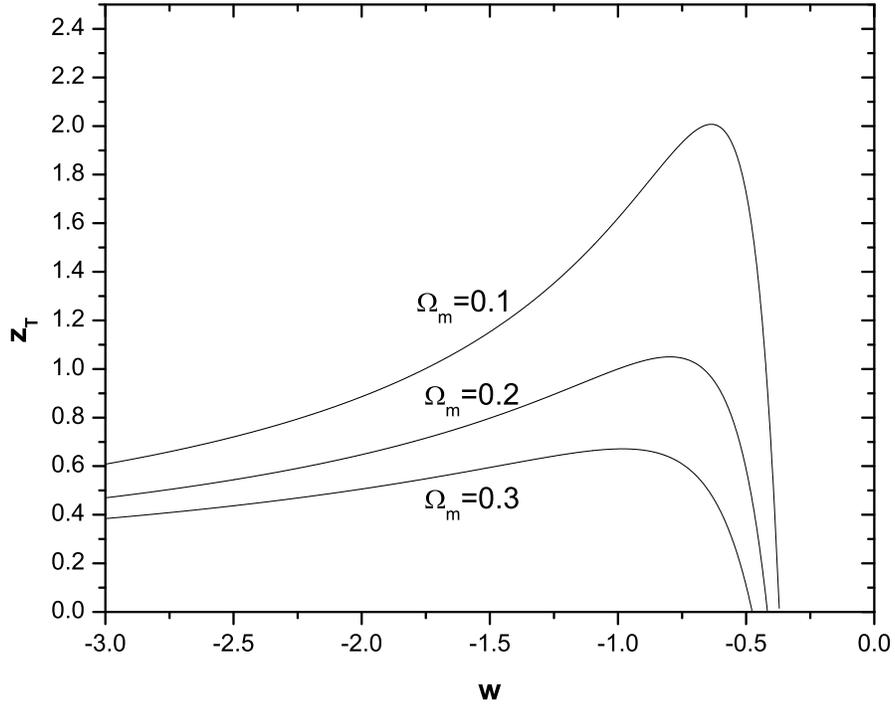}
\caption{
The transition redshift, $z_T$, at which the universe transitions
from a state of deceleration to a state of acceleration, is plotted
as a function of the equation of state of the dark energy assuming
that the equation of state of the dark energy is time-indepedent.
If the empirically determined $q(z)$ limits the transition redshift,
then bounds may be placed on the amount and redshift evolution
of the dark energy.  
}
\end{figure}

\end{document}